\documentclass[aps,prd,fleqn,superscriptaddress,floatfix,twocolumn]{revtex4}

\usepackage{graphicx,color}
\usepackage{amsmath,amssymb,amsfonts}

\newcommand{\be}{\begin{equation}}
\newcommand{\ee}{\end{equation}}
\newcommand{\ba}{\begin{eqnarray}}
\newcommand{\ea}{\end{eqnarray}}
\newcommand{\ban}{\begin{eqnarray*}}
\newcommand{\ean}{\end{eqnarray*}}
\newcommand{\nn}{\nonumber}

\newcommand{\sgn}{{\rm sgn}}

\begin{document}

\title{Charge Fluctuations from the Chiral Magnetic Effect in Nuclear Collisions}

\author{Berndt M\"uller}
\affiliation{Department of Physics \& CTMS, Duke University, Durham, NC 27708, USA}

\author{Andreas Sch\"afer}
\affiliation{Institut f\"ur Theoretische Physik, Universit\"at Regensburg,
             D-93040 Regensburg, Germany}

\date{\today}

\begin{abstract}
We derive a nonlocal effective Lagrangian for the chiral magnetic effect. An electric field is generated by winding number fluctuations of the nonabelian gauge field in the presence of a strong magnetic field.  We estimate the magnitude of charge asymmetry fluctuations with respect to the reaction plane induced by the chiral magnetic effect in relativistic heavy ion collisions to be less than $10^{-6}$, several orders of magnitude smaller than the signal observed in the STAR experiment.
\end{abstract}

\maketitle


As pointed out by Fukushima, Kharzeev, and Warringa \cite{Fukushima:2008xe}, a magnetic field induces a parallel electric current in the presence of a winding number transition in the color-SU(3) gauge field. This phenomenon, which has been called the {\em chiral magnetic effect}, can in principle be studied in off-central relativistic heavy ion collisions. Since the magnetic field is oriented parallel to the reaction plane, fluctuations of the winding number of the gauge field are expected to cause fluctuations in the electric charge distribution of emitted hadrons with respect to the reaction plane. Possible indications for such a phenomenon have been found in Au+Au collisions at the Relativistic Heavy Ion Collider, but the origin of the observed charge fluctuations among the final-state hadrons is still under debate \cite{Bzdak:2009fc}. A more reliable estimate of the magnitude of the fluctuations caused by the chiral magnetic effect would therefore be of great interest. Here we attempt such an estimate.

The electric current induced by an external magnetic field in the presence of a winding number transition can be expressed in the form
\be
\label{eq:jch}
{\bf j} = \sigma_\chi {\bf B} ,
\ee
where $\sigma_\chi$ is known as the {\em chiral magnetic conductivity} \cite{Kharzeev:2009pj}. Kharzeev {\em et al.}  gave the following expression for $\sigma_\chi$ in a thermal quark-gluon plasma \cite{Fukushima:2008xe}:
\be
\label{eq:sigma-chi}
\sigma_\chi = \sum_f \frac{3e_f^2g^2}{16\pi^4T^2}\, \int dt\, ({\bf E}^a\cdot{\bf B}^a) ,
\ee
where ${\bf E}^a, {\bf B}^a$ denote the SU(3) gauge fields, $g$ is the SU(3) gauge coupling, and $e_f$ is the electric charge of each quark flavor. We proceed by rederiving (\ref{eq:sigma-chi}) in a slightly more general context and construct an effective infrared Lagrangian encoding the chiral magnetic effect.


It is useful to first remind ourselves of the derivation of the electric analogue of eq.~(\ref{eq:sigma-chi}), i.~e.\ Ohm's law ${\bf j} = \sigma {\bf E}$ by starting from the general expression for the current density:
\be
{\bf j} = e n {\bf v} ,
\ee
where $n$ is the density of (nonrelativistic) charged particles. An external electric field accelerates the particles, and they acquire the momentum
\be
\Delta{\bf p}(t) = m {\bf v}(t) = e \int_0^t dt'\, {\bf E}(t') .
\ee
Denoting the average time between collisions that randomize the particle momentum as $t_c$, the average additional collective particle velocity is
\be 
\langle {\bf v} \rangle \approx \frac{e}{m} {\bf E} t_c .
\ee
We thus obtain for the collective current induced by the electric field:
\be
\label{eq:j}
\langle{\bf j}\rangle = e\, n\, \langle{\bf v}\rangle \approx \frac{e^2 n\, t_c}{m}\, {\bf E} 
\equiv \sigma {\bf E} ,
\ee
which is the Drude formula for the electric conductivity. In order to derive an effective action from this relation, we use the gauge ${\bf A} = -t{\bf E}$, $d{\bf A} = -t{d\bf E}$ to obtain:
\be
{\cal L}_{\rm eff} = - \int \langle{\bf j}\rangle\cdot d{\bf A} 
= t \int \sigma{\bf E}\cdot d{\bf E} = \frac{\sigma}{2} \int_0^t dt'\, {\bf E}^2 .
\ee 
The nonlocal effective Lagrangian becomes more familiar when expressed in terms of the Fourier 
transformed field ${\bf \tilde E}(\omega)$:
\be
\tilde{\cal L}_{\rm eff} =  \frac{i\sigma}{2\omega} |{\bf\tilde E}(\omega)|^2 ,
\ee 
which shows that $\sigma/\omega$ is the imaginary part of the dielectric function $\epsilon(\omega)$.

We now return to the chiral magnetic current (\ref{eq:jch}). In order to derive an expression for $\sigma_\chi$, we need to split the particle content into left-handed and right-handed components:
\be
\langle {\bf j} \rangle = e \langle n_{\rm R}{\bf v}_{\rm R} + n_{\rm L}{\bf v}_{\rm L} \rangle .
\ee
The change in the difference between left-and right-handed particle densities is governed by the axial anomaly:
\be
\frac{\partial n_5}{\partial t}  
\equiv \frac{\partial}{\partial t} \left( n_{\rm R} - n_{\rm L} \right) 
= - \frac{g^2}{8\pi^2}\, {\bf E}^a\cdot{\bf B}^a .
\ee
In order to calculate the dependence of $\langle{\bf j}\rangle$ on the magnetic field ${\bf B}$ and the excess of left-handed particles, we introduce a chemical potential $\mu_5$ for the chirality. The energy of a quark whose spin is aligned (anti-aligned) with the magnetic field in the $z$-direction is
\ba
E(p,s_z=+1/2) &=& \sqrt{p_z^2+m^2+(2n+2s_z+1)eB} 
\nonumber \\
&=& \sqrt{{\bf p}^2+m^2+2eB} ,
\nn \\
E(p,s_z=-1/2) &=& \sqrt{{\bf p}^2+m^2} .
\ea
where we have substituted the term associated to the individual Landau orbits (characterized by $n$) by ${\bf p_{\perp}}^2$ and neglected the anomalous magnetic moment contribution. We will see that only the term linear in $2eB$ contributes, which can be calculated assuming that it is infinitesimally small. This allows to substitute the discrete values of ${\bf p_{\perp}^2}$ by continuous ones.  The chiral chemical potential $\mu_5$ couples to chirality $\sigma$. For a Landau level moving with momentum $p_z$, the chirality is given by by $\sigma = 2s_z \sgn(p_3)$. Note that while ${\bf p_{\perp}}$ changes with time, the helicity of a massless quark remains unchanged due to the vector nature of the electromagnetic field.

We study the massless case in which the chiral charge density and the chiral magnetic current can be expressed as 
\ba
\label{eq:n5}
n_5 &=& \frac{1}{8\pi^3} \sum_{\sigma=\pm}\, \int d^3p\, 
\frac{\mu_5+\sigma p}{E[p,\sigma\,\sgn(p_3)] - \sigma \mu_5}
\nn \\
&&\times
\big( 1-2n_{\rm F}(E[p,\sigma\sgn(p_3)])\big)
\\
\label{eq:jz}
\langle j_z \rangle &=& \frac{e}{8\pi^3} \sum_{\sigma=\pm}\, \int d^3p\,
\frac{\sigma (\mu_5+\sigma p)}{E[p,\sigma\,\sgn(p_3)] - \sigma \mu_5} 
\nn \\
&&\times \frac{p_3}{E[p,\sigma\,\sgn(p_3)]+\sigma \mu_5}
\nn \\
&&\times
\big( 1-2n_{\rm F}(E[p,\sigma\,\sgn(p_3)])\big) ,
\ea
where $n_{\rm F}(E) = (e^{E/T}+1)^{-1}$ is the Fermi distribution. 

The leading divergences in these expression all cancel and one obtains simple finite results. The evaluation of the integrals is discussed in detail for the more difficult case $\langle {\bf j} \rangle$ in Appendix A. The results are 
\ba
n_5 &\approx& \frac{\mu_5 T^2}{3} 
\\
\langle {\bf j} \rangle &\approx& \frac{\mu_5 e^2 {\bf B}}{2\pi^2} .
\ea
We can use these two equations to eliminate $\mu_5$ and to obtain the desired relation between the chiral quark number density and the electromagnetic current:
\ba
\langle {\bf j} \rangle &\approx& \frac{3e^2{\bf B}}{2\pi^2T^2}\, n_5
\nn \\
&=& - \frac{3e^2g^2}{16\pi^4T^2}\, {\bf B}\, \int dt\, ({\bf E}^a\cdot{\bf B}^a) .
\ea
which agrees with the result obtained in ref.~\cite{Fukushima:2008xe}. Following the same argument as in the case of Ohm's law, we can now cast our result in terms of a nonlocal effective Lagrangian for the chiral magnetic effect:
\be
\label{eq:Leff-CME}
{\cal L}_{\rm eff}^{\rm (CME)}
= - \frac{e^2g^2}{8\pi^4T^2}\, \int_0^{\tau_B} dt'\, ({\bf E}\cdot{\bf B})\, \int_0^{\tau_{\rm sph}} dt''\, ({\bf E}^a\cdot{\bf B}^a) .
\ee

Different mechanisms provide three different upper limits to the time integrals in (\ref{eq:Leff-CME}): 
\begin{enumerate}
\item 
The transition time $\tau_{\rm sph}$ between gauge field configurations with different winding number; 
\item 
the time $\tau_\chi$ between many-body interactions that change the helicity of a quark; and 
\item 
the life-time $\tau_B$ of the strong magnetic field.
\end{enumerate} 
In a quark-gluon plasma or a similar highly excited QCD medium without spontaneous chiral symmetry breaking, the transition time between states of different winding number is determine by the rate of so-called {\em sphaleron} transitions. These are diffusion processes of the gauge field across the semiclassical saddle point between configurations with different winding number. The characteristic time scale for these transitions is $\tau_{\rm sph} \sim [\alpha_s^2 T \ln(1/\alpha_s)]^{-1}$ \cite{Arnold:1998cy}.

Helicity changing processes that are not related to sphaleron transitions are either governed by the nonzero current quark mass or by instantons, {\em i.~e.}\ quantum fluctuations of the winding number of the gauge field. Perturbative interactions among massless quarks and between quarks and gluons are helicity conserving. Current quark mass effects are negligible for $u$ and $d$ quarks and occur on a time scale of 1 fm/c for strange quarks. Instantons, {\em i.~e.}\ topological tunneling processes, which occur on a time scale of 0.3 fm/c in the vacuum, are strongly suppressed at nonzero temperature. The helicity decay time $\tau_\chi$ therefore is expected to be much larger than the other time scales, which are estimated to be below 1 fm/c.

The life-time of the strong magnetic field in a relativistic heavy ion collision at central rapidity is controlled by the longitudinal extent of the Lorentz contracted nuclei, which is $\tau_B \sim R/\gamma$, where $R$ is the nuclear radius and $\gamma=E_{\rm cm}/M_N$ is the Lorentz factor. Here $E_{\rm cm}$ is the beam energy per nucleon in the center-of-mass frame and $M_N$ is the nucleon mass. At the highest RHIC energy, $\gamma\approx 100$; at the Large Hadron Collider (LHC) the factor will be $\gamma\approx 3,500$. 

Even at RHIC energies, $\tau_B$ will be comparable, if not shorter than the sphaleron transition time $\tau_{\rm sph}$. This has two implications. One is that, in effect, the chiral magnetic effect will be mainly sensitive to the time averaged magnetic field pulse
\be
\overline{\bf B} = \frac{1}{\tau_B} \int_0^{\tau_B} dt\, {\bf B}(t') .
\ee
The other is that the estimate of the chiral magnetic effective action, which assumed an approximately constant magnetic field, becomes questionable. To conclude this section, we write the effective action for the chiral magnetic effect, including the contributions from all light quark flavors, as
\be
\label{eq:Leff-CME-B}
{\cal L}_{\rm eff}^{\rm (CME)}
 \approx - \frac{\sum_f e_f^2g^2\tau_B}{8\pi^4T^2}\, ({\bf E}\cdot\overline{\bf B})\, \int_0^{\tau_{\rm B}} dt''\, ({\bf E}^a\cdot{\bf B}^a) .
\ee


A local electric current induced by the magnetic field during the early stage of a relativistic heavy ion collision only becomes observable when it is reflected in the final momentum distribution of charged particles. Asakawa {\em et al.} \cite{Asakawa:2010bu} showed that the conversion of the local current into a final-state observable can occur owing to the collective outward flow of the matter. They related the charge asymmetry fluctuations with respect to the reaction plane to the vertical gradient of the expansion velocity at the time of particle freeze-out. Here we are pursuing a slightly different approach by considering the back reaction of the electric field to the chiral magnetic effect. The effective action (\ref{eq:Leff-CME-B}) is linear in the electric field strength ${\rm E}$. This means that the lowest total energy density is attained for a nonzero value of the electric field. To see this, we add (\ref{eq:Leff-CME-B}) to the free electric field energy density in the presence of the magnetic field:
\be
\label{eq:L-HE}
{\cal L}_{\rm HE}({\bf E}) = \frac{1}{2}\, f({\bf B})\, {\bf E}^2 .
\ee
where the function $f({\bf B})$ includes nonlinear quantum electrodynamic (QED) effects \cite{Mages:2010}.  By explicit calculation one finds that for a constant magnetic field of magnitude $B=(100 {\rm MeV}^2)$ these effects give rise to a very large $f({\bf B}) \approx 10$,  {\em i.~e.}\ nonlinear QED effects can reduce the induced electric field strength substantially.  Therefore, for any quantitative study one would have to calculate the strong QED effects for a realistic, time- and space-dependent magnetic field, which is a very demanding task indeed.
The minimum of the field energy is obtained for 
\be
\label{eq:Emin}
{\bf E}_{\rm min} = \frac{\sum_f e_f^2g^2\tau_B}{8\pi^4T^2f({\bf B})}\, \cdot\overline{\bf B}\, \int_0^{\tau_{\rm B}} dt''\, ({\bf E}^a\cdot{\bf B}^a)
\ee
We can also understand this relation as consequence of Maxwell's equation $\partial{\bf E}/\partial t = - {\bf j}$. The nonvanishing electric field accelerates quarks of opposite charge in opposite directions. As a result, the momentum space distributions of positive and negative charged particles will be slightly shifted with respect to each other by the momentum
\be
\Delta {\bf p} \approx q\, {\bf E}_{\rm min}\, \tau_B .
\ee
This momentum shift further deforms the thermal spectrum of the emitted particles, which is already influenced by the collective flow ${\bf v}$. The resulting spectrum
\be
E_p \frac{d^3N}{dp^3} \propto \exp\big( -\beta_f\gamma [E_p - {\bf v}\cdot({\bf p} - \Delta{\bf p})] \big) ,
\ee
where $\beta_f=T_f^{-1}$ is the inverse freeze-out temperature and $\gamma = (1-v^2)^{-1/2}$ is the Lorentz factor associated with the collective flow, implies a charge ratio (with $q=\pm e$):
\be
\frac{dN^{(+)}}{dN^{(-)}} \approx \exp\big( 2 e  \tau_B {\bf v}\cdot{\bf E}_{\rm min}/T_f \big) .
\ee
The sign of ${\bf v}\cdot{\bf E}_{\rm min}$ fluctuates randomly between different coherence domains of the Chern-Simons number density ${\bf E}^a\cdot{\bf B}^a$. The total charge asymmetry is thus given by
\be
\label{eq:N+N-}
\Delta^\pm = 
\frac{dN^{(+)}-dN^{(-)}}{dN^{(+)}+dN^{(-)}} \approx  \frac{ e\,  \tau_B\, {\bf v}\cdot{\bf E}_{\rm min}}{T_f \sqrt{N}} ,
\ee
where $N$ denotes the number of independent domains. 

This formula assumes that the momentum kick received by the quarks during the action of the chiral magnetic effect survives until freeze-out. One would expect that diffusive processes in the quark-gluon plasma result in a reduction of the initially generated momentum difference between quarks with different charges. On the other hand, the small initial spatial separation between positive and negatively charged quarks may amplify the final momentum shift owing to the gradient of the collective velocity field in the direction perpendicular to the reaction plane, as discussed in ref.~\cite{Asakawa:2010bu}.


We now estimate the magnitude of the expected fluctuating charge asymmetry. Using the result for the time integrated magnetic field in Au+Au collisions given in the Appendix of ref.~\cite{Asakawa:2010bu}, we have
\be
\tau_B |e\overline{\bf B}| \approx \frac{2.3\, Z\alpha\, b}{R^2} ,
\ee
where $R$, and $Z$ are the nuclear radius and charge, respectively, and $b$ is the impact parameter. Summing over up, down, and strange quarks and neglecting the quark masses, we find
\be
|e{\bf E}_{\rm min}| \approx \frac{8\, \alpha\, 2.3\, Z\alpha\, b}{\pi\, f({\bf B}) T^2\, R^2}\, \frac{|Q_5|}{V} ,
\ee
where $|Q_5|/V$ denotes the volume density of winding number transitions. Lacking a more reliable estimate, we assume one winding number transition per volume $\bar{\rho}^3$ with $\bar{\rho} \approx 0.5~{\rm fm}$ during the presence of the strong magnetic field. Finally, we choose $T\approx 350$ MeV as a proxy for the quark density during the strongly magnetized phase of the heavy ion collision, and set $f({\bf B})\approx 10$. For $Z=79$ and $R=7$ fm this yields
\be 
|e{\bf E}_{\rm min}| \approx (6~{\rm MeV})^2\, (b/R) .
\ee
Finally, using the estimate $\tau_B \approx 2R/\gamma_{\rm cm}$ with $\gamma_{\rm cm}=100$ for the duration of the strong magnetic field, a final collective flow velocity $|{\bf v}| \approx 0.5c$, a freeze-out temperature $T_f\approx 150$ MeV, and $N=\pi R^2/\bar{\rho}^2 \approx 600$ independent winding number domains, we obtain
\be
\Delta^\pm \approx 3.5\times 10^{-7} \frac{b}{R} ,
\ee
which is much smaller than the charge correlation observed in the STAR experiment \cite{Abelev:2009uh,Abelev:2009tx}.

This estimate can be improved in many ways. The most important improvement would be a more reliable estimate of the winding number fluctuation density during the strongly magnetized phase of the heavy ion reaction. This requires calculations of the contribution to $Q_5/V$ from the interaction between the  color glass condensate clouds contained in the two colliding nuclei \cite{Asakawa:2010bu} and of the sphaleron transition rate in pre-equilibrated quark-gluon plasma \cite{Kharzeev:2001ev}. Another approach to this problem involve lattice QCD simulations of the winding number fluctuation rate in the presence of a strong external magnetic field \cite{Buividovich:2009wi,Buividovich:2009zj,Abramczyk:2009gb}. 

A second possible improvement concern the calculation of the electric current induced by a winding number transition in the magnetic field, since the quark distribution is unlikely to be thermalized during the early period when the magnetic field is strong. One way in which this estimate can be improved is by calculating the quark-antiquark distribution liberated by the interacting color glass condensates \cite{Gelis:2005pb}. 

Another important improvement would be the inclusion of charge transport processes during the hydrodynamic expansion of the quark-gluon plasma and the hadronic freeze-out phase. Such calculations would facilitate more realistic estimates of the final charge asymmetry fluctuations with respect to the reaction plane. 

Finally, the nonlinear QED effects have to be calculated for a more realistic field configuration. Although is is not relevant for the scenario of ``small'' winding number coherence domains considered here, we mention a QED effect that may limit the chiral magnetic effect for large coherence domains. Electron-positron pair production limits the strength of stable electric fields to $eE \leq eE_{\rm max} = 2m_e/ \bar{\rho}$, where $m_e$ is the electron mass and $\bar{\rho}$ denotes the winding number domain size.

In summary, we have derived a nonlocal effective Lagrangian for the chiral magnetic effect. The charge current corresponding to this Lagrangian agrees with earlier calculations. We show that a fluctuating electric field is generated by winding number fluctuations of the nonabelian gauge field in the presence of a strong magnetic field, and we estimated its effect on the charge asymmetry of the spectra of emitted hadrons. We estimated the fluctuating relative charge particle asymmetry in the final state and identified several opportunities for future improvements of the estimate.

\begin{acknowledgments}  
We acknowledge helpful discussions  with K.~Fukushima and A.~Majumder. We thank C.~Athanasiou and K.~Rajagopal for pointing out an error in an earlier version of our manuscript. This work was supported in part by grants from the U.S. Department of Energy (DE-FG02-05ER41367) and from the BMBF. The research reported here was initiated during the HESI10 Workshop at the Yukawa Institute of Kyoto University.
\end{acknowledgments}
\bigskip

\appendix
\section{Chiral electric current}

In order to calculate the dependence of $\langle{\bf j}\rangle$ on the magnetic field ${\bf B}$ and on the excess of right-handed particles, we follow ref.~\cite{Fukushima:2008xe} and introduce a chemical potential $\mu_5 \ll T$ for the chirality $\sigma=\pm 1$. The energy of a quark whose spin is aligned (anti-aligned) with the magnetic field is
\ba
E(p,s_z) &=& \sqrt{p_z^2+m^2+(2n+2s_z+1)eB} 
\nn \\
&=& \sqrt{{\bf p}^2+m^2+(s+1)eB}
\nn \\
&\equiv& E_{p,s}  .
\ea
where $s\equiv 2s_z=\pm 1$ is twice the eigenvalue of the spin component along the magnetic field, here taken as the $z$-direction. The eigenvalues of the transverse kinetic energy are given by the Landau levels $\langle {\bf p}_\perp^2\rangle_n = 2n eB$. We will assume in the following that the magnetic fields is weak ($eB \ll T^2$) so that the Landau orbits form a quasi-continuum and the sum over $n$ can be approximated by an integral over ${\bf p}_\perp$. 

The interaction of the quark with the magnetic field does not violate chirality. This implies the relation $\sigma=s\, {\rm sgn}(p_z)$. We thus obtain the following distribution functions for right- and left-handed quarks in the presence of a magnetic field:
\be
\label{eq:dist}
n_\sigma({\bf p}) = n_{\rm F}(E_{p,s} - \mu_5 \sigma)  ,
\ee
where $n_{\rm F}(E)$ denotes the Fermi-Dirac distribution. We thus find the following expressions for the net chirality and the average velocity density in the presence of the magnetic field:
\be
\label{eq:chirality}
j_z = -e \sum_s \int\frac{d^3p}{(2\pi)^3} \frac{p_z}{E_{p,s}} \tanh\frac{E_{p,s} - \mu_5 \sigma}{2T}  .
\ee
We now expand for small $\mu_5$ and $B$ up to linear order in $\mu_5$ and the magnetic field $B$:
\begin{widetext}
\ba
j_z &\approx& \mu_5 e(eB) \sum_s \int\frac{d^3p}{(2\pi)^3} p_z \frac{\partial E_{p,s}}{\partial (eB)}
  \frac{\partial}{\partial E_{p,s}} \left( \frac{1}{E_{p,s}} \frac{\partial}{\partial E_{p,s}}
  \tanh\frac{E_{p,s} - \mu_5 \sigma}{2T} \right)_{\mu_5=B=0}
\nn \\
&=& - \mu_5 e(eB) \int\frac{d^3p}{(2\pi)^3}\, |p_z|  \frac{1}{E_p}
  \frac{\partial}{\partial E_{p}} \left( \frac{1}{E_{p}} \frac{\partial}{\partial E_{p}}
  \tanh\frac{E_{p}}{2T} \right) ,
\ea
\end{widetext}
where the sum over chiralities has contributed a factor 2, and $E_p^2 = {\bf p}^2+m^2$. In the following we will set $m=0$. As $(s+1)eB$ contributes only for $s=+1=\sigma\, \sgn(p_z)$, the integral over $p_z$ only runs over positive (negative) $p_z$ for positive (negative) chiralities $\sigma$. Performing the angular integrals and setting $E_p=p=|{\bf p}|$ for massless quarks, we obtain
\ba
j_z &=& - \frac{\mu_5 e(eB)}{8\pi^2} \int dp\, p^2\, \frac{\partial}{\partial p} \left( \frac{1}{p} \frac{\partial}{\partial p}
  \tanh\frac{p}{2T} \right) 
\nn \\
&=& - \frac{\mu_5 e(eB)}{8\pi^2} \int dx\, x^2\, \frac{\partial}{\partial x} \left( \frac{1}{x \cosh^2x} \right) 
\nn \\
&=& \frac{\mu_5 e(eB)}{2\pi^2} .
\ea
This agrees with the result obtained by Fukushima {\em et al.} \cite{Fukushima:2008xe}.

\end{document}